\documentclass[sigplan,10pt]{acmart}
\renewcommand\footnotetextcopyrightpermission[1]{}

\usepackage{amsmath}
\usepackage{algorithmic}
\usepackage{graphicx}
\usepackage{textcomp}
\usepackage{xcolor}
\usepackage{url}
\usepackage{graphicx}
\usepackage{amsmath}
\usepackage{bm}
\usepackage{comment}
\usepackage{array}
\usepackage{xspace}
\usepackage{psfrag}
\usepackage{subcaption}
\usepackage[linesnumbered,ruled,vlined]{algorithm2e}
\usepackage{alltt}
\usepackage{multirow}
\usepackage{array}
\usepackage{rotating}
\usepackage{datetime}
\usepackage{amsthm}
\usepackage{epsfig}
\usepackage{endnotes}
\usepackage{epstopdf}
\usepackage{booktabs}
\usepackage{setspace}
\usepackage{color}
\usepackage[shortlabels]{enumitem}
\usepackage{float}
\usepackage{makecell}
\usepackage{pifont}
\usepackage{textcomp}
\usepackage{tabularx}

\theoremstyle{definition}

\theoremstyle{remark}

\newcommand{\sysname}{\textit{AdaptCore}\xspace}

\SetKwInput{KwInput}{Input}                
\SetKwInput{KwOutput}{Output}              
\newcommand{\RNum}[1]{\uppercase\expandafter{\romannumeral #1\relax}}

\newcommand{\squishlist}{
  \begin{list}{$\bullet$}
    { \setlength{\itemsep}{0pt}      \setlength{\parsep}{3pt}
      \setlength{\topsep}{3pt}       \setlength{\partopsep}{0pt}
      \setlength{\leftmargin}{3.5mm} \setlength{\labelwidth}{1em}
      \setlength{\labelsep}{0.5em} } }

    \newcommand{\squishend}{
  \end{list}  }

\newcounter{packednmbr}

\newcommand{\head}[1]{\noindent\textbf{#1}}

\usepackage{listings}
\usepackage{xcolor}

\definecolor{codegreen}{rgb}{0,0.6,0}
\definecolor{codegray}{rgb}{0.5,0.5,0.5}
\definecolor{codepurple}{rgb}{0.58,0,0.82}
\definecolor{backcolour}{rgb}{0.96,0.96,0.96}

\lstdefinestyle{ascendstyle}{
    backgroundcolor=\color{backcolour},   
    commentstyle=\color{codegreen}\itshape,
    keywordstyle=\color{blue}\bfseries,
    numberstyle=\tiny\color{codegray},
    stringstyle=\color{codepurple},
    basicstyle=\ttfamily\scriptsize, 
    breakatwhitespace=false,         
    breaklines=true,                 
    captionpos=b,                    
    keepspaces=true,                 
    numbers=left,                    
    numbersep=5pt,                  
    showspaces=false,                
    showstringspaces=false,
    showtabs=false,                  
    tabsize=2,
    escapeinside={(*@}{@*)} 
}
\begin{document}

\title{Adaptive Matrix Multiplication for Dynamic Shapes on Ascend NPUs}

\author{Yuhang Zhou\normalfont\textsuperscript{*}}
\affiliation{%
  \institution{Nanjing University}
  \department{State Key Laboratory for Novel Software Technology}
  \city{Nanjing}
  \country{China}
}

\author{Peng Jiang\normalfont\textsuperscript{*}}
\affiliation{%
  \institution{Nanjing University}
  \department{State Key Laboratory for Novel Software Technology}
  \city{Nanjing}
  \country{China}
}

\author{Qianyu Jiang}
\affiliation{%
  \institution{Nanjing University}
  \department{State Key Laboratory for Novel Software Technology}
  \city{Nanjing}
  \country{China}
}

\author{Zhibin Wang}
\affiliation{%
  \institution{Nanjing University}
  \department{State Key Laboratory for Novel Software Technology}
  \city{Nanjing}
  \country{China}
}

\author{Xinghui Tian}
\affiliation{%
  \institution{Huawei}
  \country{China}
}

\author{Jianwei Zhou}
\affiliation{%
  \institution{Huawei}
  \country{China}
}

\author{Songxiang Zhu}
\affiliation{%
  \institution{Huawei}
  \country{China}
}

\author{Jingyi Zhang}
\affiliation{%
  \institution{Huawei}
  \country{China}
}

\author{Junsong Wang}
\affiliation{%
  \institution{Huawei}
  \country{China}
}

\author{Chen Tian}
\affiliation{%
  \institution{Nanjing University}
  \department{State Key Laboratory for Novel Software Technology}
  \city{Nanjing}
  \country{China}
}

\begin{abstract}
Matrix Multiplication (MatMul) faces a ``generalization crisis'' driven by highly dynamic tensor shapes. This crisis is particularly acute on Ascend NPUs, where explicitly controlled architectures and strict physical constraints render existing GPU-centric optimizations ineffective. To resolve this, we propose \sysname, an adaptive framework for universally high-performance MatMul on Ascend NPUs. \sysname systematically decouples operator optimization into spatial tiling and instruction orchestration. It first maps dynamic shapes into a hardware-aware 2D tiling taxonomy to balance on-chip capacity limit and multi-core parallelism. Furthermore, it integrates a composable optimization library with a deterministic analytical performance model. By mathematically evaluating hardware state mutations, \sysname proactively selects and caches optimal implementations, enabling $O(1)$ overhead runtime dispatching. Evaluations demonstrate that \sysname delivers a remarkable 1.85$\times$ mean speedup across 80,000 input shapes, and achieves up to a 1.48$\times$ acceleration in representative end-to-end models over the highly-tuned native vendor library (ACLNN).
\end{abstract}


\maketitle
\pagestyle{plain} 

\makeatletter
\if@ACM@anonymous\else
  \newcounter{savefootnote}
  \setcounter{savefootnote}{\value{footnote}}
  \renewcommand{\thefootnote}{*}
  \footnotetext{These authors contributed equally.}
  \setcounter{footnote}{\value{savefootnote}}
\fi
\makeatother

\section{Introduction}
\label{sec:intro}
Matrix Multiplication (MatMul) has long been the computational core of AI workloads, highly optimized for static uniform input shapes. However, as workloads become increasingly complex, MatMul operators face a severe ``generalization crisis.'' On the one hand, input dimensions exhibit massive variance, often spanning multiple orders of magnitude (e.g., from 1 to 100,000 in recommendation scenarios~\cite{naumov2019deeplearningrecommendationmodel,10.1145/3523227.3547405, 10.1145/3579355}), degrading from standard square matrices to highly skewed matrices. On the other hand, emerging model architectures introduce unpredictable fluctuations to input dimensions, such as the real-time gating mechanism in Mixture-of-Experts (MoE) models~\cite{fedus2022switchtransformersscalingtrillion, hwang2023tuteladaptivemixtureofexpertsscale}. Consequently, traditional operators optimized for specific static shapes inevitably suffer a sharp decline in performance.

This dynamic shape crisis is particularly severe on Ascend NPUs. Unlike GPUs, which follow the SIMT paradigm and hardware-managed hierarchical memory, Ascend NPUs~\cite{ascend_whitepaper,Liao2021Ascend} feature an explicitly managed SIMD architecture coupled with decoupled, complex memory buffers (e.g., L1 and L0 A/B/C). This requires developers to manually control instruction sequences and dependencies to ensure pipeline efficiency. Constrained by strict hardware limits and explicit control requirements, MatMul operators easily fall into a dilemma when encountering irregular dynamic shapes: either the tiles are too small, leaving compute units idle, or the tiles are too large, exceeding buffer capacity and triggering massive memory bandwidth pressure. 

Unfortunately, existing operator optimizations fail to achieve both architectural awareness and runtime flexibility. High-level auto-tuning frameworks (e.g., Triton~\cite{10.1145/3315508.3329973}, cuTile~\cite{cutile_python}) significantly lower the programming barrier but are intrinsically designed for GPUs. When ported to Ascend NPUs, they often ignore hardware characteristics, yielding sub-optimal performance and introducing search overhead. Similarly, while static template libraries (e.g., CUTLASS~\cite{cutlass_github}) and specialized JIT compilers (e.g., DeepGEMM~\cite{deepgemm_github}) maximize hardware utilization, they either fail to handle dynamic shapes gracefully or strictly target specific hardware (e.g., Hopper).

In practice, developing such an ideal MatMul operator for dynamic shapes on Ascend NPUs is hindered by two challenges: \textit{(i) Hardware-Constrained Tiling.} Mapping massive dynamic workloads to optimal tile sizes requires a balance between on-chip buffer capacities and multi-core parallelism. Enlarging tiles to save memory bandwidth inevitably risks buffer overflow, while shrinking them leaves compute cores starved. \textit{(ii) Coupled Optimization Space.} Beyond tiling, fine-grained instruction optimizations (e.g., memory padding, preloading) introduce complex trade-offs. Each technique accelerates execution but also incurs inherent overheads. Without a method to quantitatively evaluate these combinatorial impacts, identifying the optimal implementation remains an open challenge.

To overcome these challenges, we propose \sysname, an adaptive framework designed for universally high-performance MatMul execution on Ascend NPUs. At its core, \sysname is driven by the insight that operator optimization can be decoupled into macro-level spatial tiling and micro-level instruction orchestration. 
(i) To resolve the tiling conflict between memory constraints and load balance, \sysname abandons empirical heuristics. Instead, it maps dynamic input shapes into a rigorous hardware-aware 2D taxonomy, classifying them into four base tiling templates to achieve the balance between buffer capacity and multi-core parallelism. 
(ii) To navigate the complex optimization space, \sysname constructs a unified analytical performance model based on Ascend's deterministic instruction pipeline. By mathematically evaluating the impact of computation, bandwidth, and pipeline optimizations, this model identifies the global optimal implementation across various tiling and optimization combinations. 
Finally, to seamlessly fuse these two stages for dynamic workloads, \sysname adopts a synergistic offline-online execution mechanism. In the offline phase, the analytical model proactively explores, compiles, and caches the optimal implementations into a lookup table. During online execution, \sysname achieves $O(1)$ overhead runtime dispatching through rapid shape indexing.

In summary, this paper makes the following contributions:
\begin{itemize}
    \item We formulate MatMul tiling as a hardware-aware 2D template space defined by spatial task sufficiency and the accumulation regime, associating each shape with a template and hierarchical tiling parameters.

    \item We construct an analytical model and a composable optimization library to select legal optimization configurations by their data movement, resource efficiency, and pipeline schedule.

    \item We design an offline-selection/online-dispatch workflow that encodes each preselected implementation in a compact \texttt{TilingKey}, enabling constant-time runtime kernel lookup without online exploration.

    \item We evaluate \sysname on 80{,}000 dynamic shapes and five recommendation models. It achieves a $1.85\times$ mean single-operator speedup over vendor library ACLNN and delivers $1.09\times$--$1.48\times$ end-to-end speedups.
\end{itemize}

The rest of this paper is organized as follows. Section~\ref{sec:background} provides the background, and Section~\ref{sec:overview} presents the system overview. Section~\ref{sec:tiling} introduces the hardware-aware tiling taxonomy. Section~\ref{sec:optimization} presents model-guided selection, analytical performance modeling, and the optimization library. Section~\ref{sec:evaluation} reports the evaluation. Section \ref{sec:limitations} discusses limitations, and Section \ref{sec:conclusion} concludes.
\section{Background}\label{sec:background}

\subsection{Ascend Architecture and Physical Constraints}
\begin{figure}[tbp]
    \centering
    \includegraphics[width=0.9\linewidth]{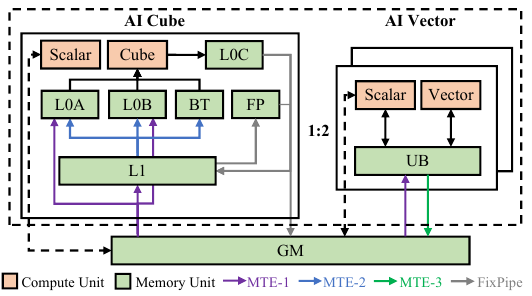}
    \caption{Architectural overview of the Ascend 910 NPU.}
    \label{fig:ascend}
\end{figure}

To fully understand the necessity of specialized operator optimization on the Ascend architecture, we first contrast its hardware design with the dominant GPU paradigm.

\textbf{GPU: Implicit Synchronization.} 
Modern GPUs, such as NVIDIA's Blackwell architecture, employ a Single Instruction Multiple Threads (SIMT) execution model and hierarchical memory. Crucially, GPUs utilize hardware mechanisms to help developers hide memory latency and manage data flow, thereby reducing the need for explicit instruction pipeline orchestration. For example, the Tensor Memory Accelerator (TMA) autonomously fetches matrix tiles from Global Memory to Shared Memory, while hardware-level barriers (e.g., \texttt{mbarrier}) implicitly resolve dependencies.

\textbf{Ascend NPU: Explicit Execution.} 
Unlike GPUs, the Ascend NPU~\cite{Liao2021Ascend} features a deeply decoupled Single Instruction Multiple Data (SIMD) architecture. As illustrated in Figure~\ref{fig:ascend}, the computational core (AICore) is composed of asynchronous units: the Cube unit (matrix multiplication), the Vector unit (element-wise), multiple Memory Transfer Engines (MTEs), and complex on-chip memory buffers, including the L1 Buffer, L0A/B/C Buffers and Unified Buffer (UB). To unlock peak performance, developers must explicitly orchestrate both the data movement across a complex on-chip memory hierarchy and the instruction synchronization.

As shown in Listing~\ref{lst:ascend_code}, the MatMul operator on Ascend consists of the following instructions: (i) The MTE2 transfers data from Global Memory (GM) to L1 Buffer. (ii) The MTE1 routes the left matrix A from L1 Buffer to L0A, and the right matrix B to L0B. (iii) The Cube unit reads operands from L0A/B to perform matrix multiplication, accumulating the results in L0C. (iv) The FixPipe (handles format conversion and post-processing) writes the results back to GM. 

\begin{lstlisting}[language=C++, caption={Abstracted Ascend C MatMul kernel}, label={lst:ascend_code}]
GlobalTensor<half> gm_A, gm_B; GlobalTensor<float> gm_C; 
TQue<QuePosition::VECIN, 1> qL1_A, qL1_B;
TQue<QuePosition::A1, 1> qL0A;
TQue<QuePosition::B1, 1> qL0B;
TQue<QuePosition::CO1, 1> qL0C;
/* (i) GM -> L1 (MTE2) */
DataCopy(l1_A, gm_A, TSIZE); DataCopy(l1_B, gm_B, TSIZE);
qL1_A.EnQue(l1_A); qL1_B.EnQue(l1_B); // SetFlag(L1)
/* (ii) L1 -> L0A/L0B (MTE1) */
l1_A = qL1_A.DeQue<half>(); l1_B = qL1_B.DeQue<half>(); // WaitFlag(L1) 
DataCopy(l0a, l1_A, TSIZE); DataCopy(l0b, l1_B, TSIZE); 
qL0A.EnQue(l0a); qL0B.EnQue(l0b); // SetFlag(L0A/B)
/* (iii) L0A, L0B -> Cube -> L0C */
l0a = qL0A.DeQue<half>(); l0b = qL0B.DeQue<half>(); // WaitFlag(L0A/B)
Mmad(l0c, l0a, l0b, ...);            
qL0C.EnQue(l0c); // SetFlag(L0C)
/* (iv) L0C -> GM (FixPipe) */
l0c = qL0C.DeQue<float>(); // WaitFlag(L0C)  
DataCopy(gm_C, l0c, TSIZE);
\end{lstlisting}

In fact, the performance of MatMul on Ascend NPUs is fundamentally impacted by two constraints:
\textit{(i) Limited Buffer Capacities.} The capacities of the dedicated on-chip buffers directly impact the efficiency of operand fetching (L0A/B) and result accumulation (L0C). In particular, if the intermediate partial sums overflow the L0C buffer, the AICore is forced to flush data back to the off-chip GM for accumulation, severely degrading the memory bandwidth.
\textit{(ii) Instruction Pipeline Efficiency.} Overall execution throughput relies heavily on how developers orchestrate instruction sequences and synchronization dependencies. For example, through fine-grained instruction optimization, we can overlap the MTE data fetching for the next block with the ongoing Cube computation to improve efficiency. Conversely, sequential execution results in poor performance.

\subsection{The Dynamic Shape Crisis of MatMul}\label{sec:generalization_challenge}


In modern AI workloads, Matrix Multiplication is no longer confined to the static, uniform input dimensions characteristic of traditional dense models. Instead, it faces two profound generalization challenges:
\textit{(i) Extreme Dimensional Variance.} In industrial-scale recommendation systems, the input dimensions ($M, N, K$) of MatMul operators exhibit massive variance. The shape sizes can span a magnitude of over $10^5$, ranging from standard square matrices to highly skewed ``Tall-Skinny'' or ``Short-Wide'' matrices. 
\textit{(ii) Real-Time Dynamic Fluctuations.} Furthermore, in emerging Mixture-of-Experts (MoE) architectures, sparse activation mechanisms dynamically route tokens to a subset of experts. This real-time decision-making causes the underlying GEMM input dimensions to fluctuate unpredictably on the fly.

\begin{figure}[tbp]
    \centering
    \includegraphics[width=0.75\linewidth]{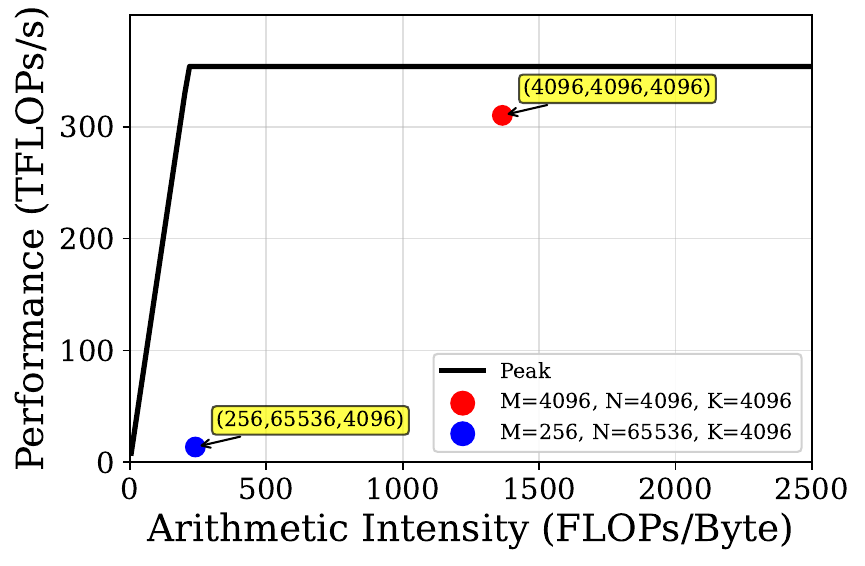}
    \caption{Roofline analysis: square vs. skewed MatMul shapes on Ascend 910.}
    \label{fig:dynamic}
\end{figure}

Consequently, operators heavily optimized for specific shapes suffer severe performance degradation when confronted with these dynamic variations. Based on the roofline analysis on Ascend NPUs in Figure~\ref{fig:dynamic}, for a standard square matrix (e.g., $M=4096, N=4096, K=4096$), a fixed tiling strategy ($256 \times 256$) can perfectly balance the workload between compute units and memory units, allowing the hardware to approach its peak performance. 
However, when the same strategy is applied to highly irregular shapes (e.g., $M=256, N=64K, K=4096$), it triggers a dual-sided hardware crisis. On the compute side, the constrained $M$ dimension drastically reduces the computational workload per tile, leaving the Cube constantly starved for data. Concurrently, on the memory side, fetching static tiles from a large $N$ dimension ($64K$) forces the DMA engine into inefficient strided memory accesses. This causes both compute throughput and memory bandwidth to be far below the hardware limits.

\subsection{Existing Work and Limitations}
\label{sec:existing}
To resolve the aforementioned crisis, an ideal Ascend MatMul operator must possess both deep architectural awareness and runtime flexibility. However, existing solutions fall into a Catch-22 dilemma, failing to achieve this goal.

\textbf{High-Level Compilers and Auto-Tuning.} OpenAI's Triton~\cite{10.1145/3315508.3329973} and NVIDIA's cuTile~\cite{cutile_python} use auto-tuners to search for optimal block-level parallelism. While highly productive, they are intrinsically tailored to the implicit synchronization of GPUs. When ported to explicitly controlled Ascend NPUs, they exhibit severe ``architectural blindness,'' failing to guide an efficient instruction pipeline. Furthermore, auto-tuning may incur prohibitive search overheads when faced with the dynamic shapes in practical workloads.

\textbf{Static Libraries and Specialized Compilers.} Conversely, expert-level template libraries prioritize peak hardware utilization but sacrifice adaptability. For example, NVIDIA's CUTLASS~\cite{cutlass_github} delivers near-handwritten performance via C++ templates, yet its rigid static configurations fail to gracefully handle extreme dynamic shapes (e.g., ``Tall-Skinny'' matrices). While specialized compilers like DeepSeek's DeepGEMM~\cite{deepgemm_github} solve this via Just-In-Time heuristics, they are locked into the specific hardware architecture (NVIDIA Hopper), offering zero portability to Ascend NPUs.

Fortunately, the native CATLASS template library~\cite{catlass_gitcode} successfully abstracts Ascend's explicit memory routing and synchronization barriers into reusable C++ components. Rather than reinventing these intricate low-level instructions from scratch, utilizing CATLASS guarantees hardware affinity. However, it is fundamentally static. Fully unlocking its potential under dynamic irregular workloads is an challenge.

\section{Overview}\label{sec:overview}



In this section, we first summarize core challenges in optimizing the performance of MatMul operators on Ascend NPUs, and then introduce our framework \sysname.
\subsection{Core Challenges}\label{sec:challenges}
In fact, optimizing the MatMul operator on Ascend NPUs can be distilled into solving two deeply intertwined challenges:

\textbf{Challenge 1: How to select optimal tiling configurations under Ascend hardware constraints?} 
The Ascend architecture imposes rigorous physical limitations, particularly the limited on-chip buffers (e.g., L0C) and complex coordination across multiple cores. The essence of the tiling problem lies in mapping dynamic input shapes to appropriate tile sizes. However, enlarging the tiles to save memory bandwidth inevitably breaches the on-chip memory limit, while shrinking them leads to low multi-core utilization. Given the variance in input shapes, finding the optimal tiling policy is extremely difficult.

\textbf{Challenge 2: How to navigate the highly coupled, multi-dimensional optimization space?} 
Beyond tiling, maximizing MatMul performance relies on fine-grained software optimizations, such as scalar elimination, memory padding, and preloading. These optimizations span multiple dimensions, including compute efficiency, memory bandwidth, and pipeline overlap. However, each optimization inevitably introduces inherent overheads alongside its performance gain. For example, preloading exacerbates on-chip buffer occupancy, while memory padding incurs extra format-conversion computation overhead. Composing multiple optimizations makes these performance impacts even more complex. The lack of quantitative evaluation of operator performance makes it difficult to determine an efficient operator implementation.

\begin{figure}[tbp]
    \centering
    \includegraphics[width=\linewidth]{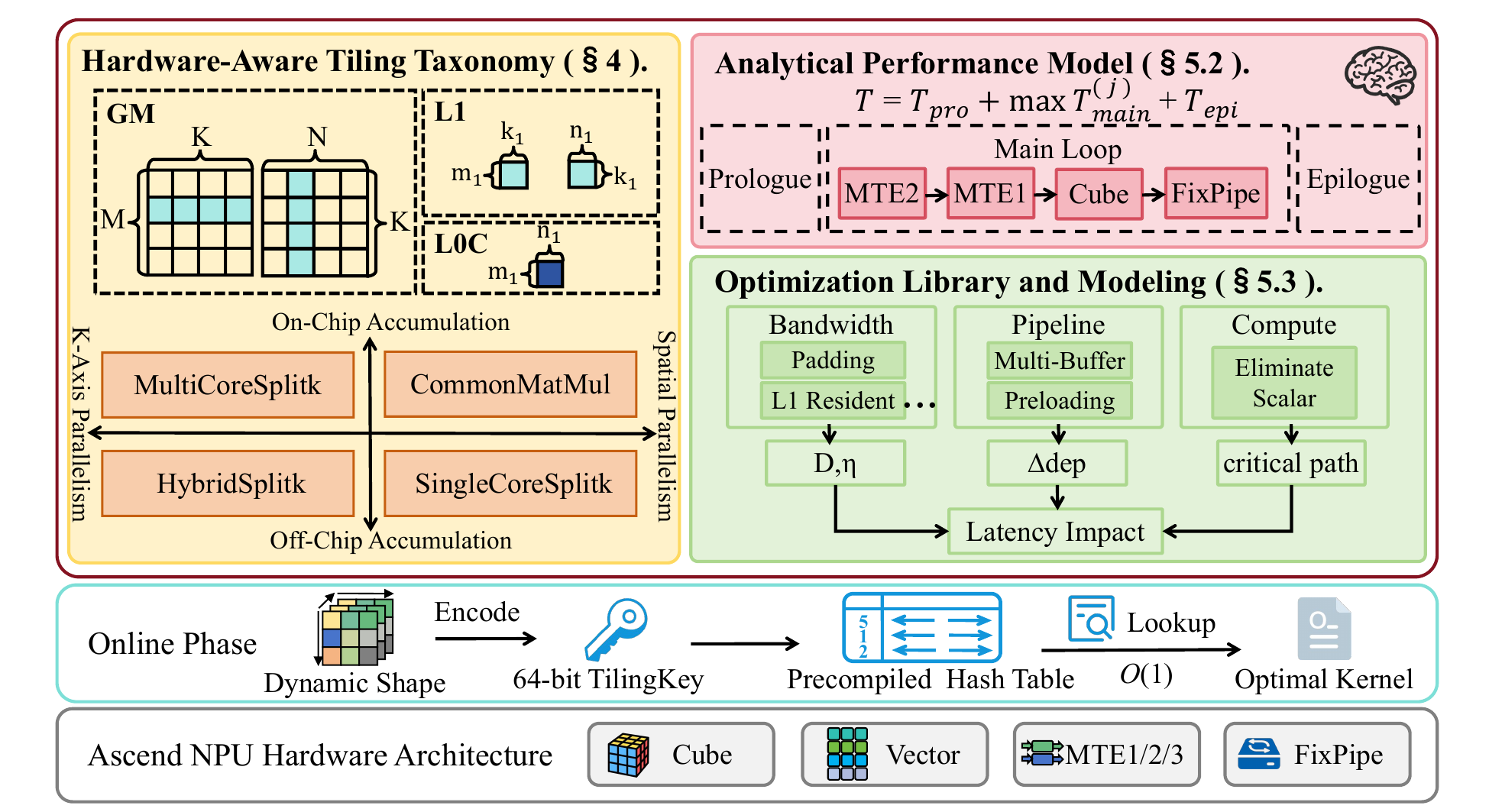}
    \caption{System overview of the proposed adaptive MatMul framework.}
    \label{fig:architecture}
\end{figure}

\subsection{System Architecture}\label{sec:architecture}
As shown in Figure~\ref{fig:architecture}, \sysname consists of three components that collaborate within a unified offline-online workflow.

\textbf{Hardware-Aware Tiling Taxonomy (\S\ref{sec:tiling}).} 
\sysname systematically maps dynamic input shapes into a 2D decision space governed by Ascend hardware constraints. This taxonomy is constructed along two orthogonal dimensions: spatial parallelism (multi-core task demands) and data reuse (the L0C capacity). The two axes naturally divide the tiling space into four distinct quadrants, each representing a specific tiling template. For example, shapes with small $M$ and $N$ map to \texttt{MultiCoreSplitK}, splitting the $K$-axis to keep all AICores active. Conversely, massive read-bound shapes map to \texttt{SingleCoreSplitK}, which enlarges tiles to maximize data reuse while overflowing the L0C buffer.

\textbf{Analytical Performance Model (\S\ref{sec:model}).} 
\sysname proposes an analytical performance model built on two abstractions of Ascend's explicit execution: (i) a three-phase decomposition of every MatMul execution into \textit{Prologue}, \textit{Main Loop}, and \textit{Epilogue}; and (ii) a four-stage instruction pipeline (MTE2 $\rightarrow$ MTE1 $\rightarrow$ Cube $\rightarrow$ FixPipe) within each Main Loop iteration. Under this view, the per-iteration latency is precisely the critical-path length of this pipeline, which the model evaluates in closed form. Based on this, \sysname provides a unified analytical framework, enabling microsecond-level latency prediction without runtime profiling.

\textbf{Optimization Library and Modeling (\S\ref{sec:library}).}
Considering the vast optimization space, as shown in Table~\ref{tab:optimizations}, \sysname componentizes representative optimizations into a library and categorizes them by the bottlenecks they address: (i) \textit{Bandwidth optimizations} satisfy strict alignment constraints and manage memory hierarchies to saturate data read/write throughput. (ii) \textit{Pipeline optimizations} manipulate the instruction streams to maximize the execution overlap between asynchronous compute and memory units. (iii) \textit{Compute optimizations} eliminate redundant control logic to ensure dense computing operations dominate the critical path. By feeding each module into the analytical model, \sysname quantitatively predicts its latency impact.

In practical application, \sysname comprises two distinct execution phases.
During the \textit{offline phase}, \sysname injects the composable optimization modules into specific templates, utilizes the performance model to evaluate all valid combinations, and registers the optimal kernel into a pre-compiled hash table named \texttt{launch\_map}. 
During the \textit{online phase}, \sysname encodes the incoming shape into a compact 64-bit \texttt{TilingKey}. This key serves as a direct index to perform an $O(1)$ lookup within the \texttt{launch\_map}, retrieving the optimal implementation with near-zero runtime overhead.                                    
\section{Hardware-Aware Tiling Taxonomy}\label{sec:tiling}
%

In this section, we first introduce the fundamental tiling principles for the MatMul operator on the Ascend NPU and systematically categorize diverse tiling scenarios into specific templates. We then present several case studies to demonstrate the superiority of tiling templates.

\subsection{Rethinking Tiling on Ascend NPUs}
Unlike GPUs that rely on SIMT thread schedulers to implicitly hide memory latencies, Ascend NPUs employ a decoupled SIMD design comprising independent MTE, Cube, and Vector units. Therefore, hardware synchronization, pipeline overlapping, and data transfer are not automatically managed, but explicitly governed by the tiling strategy. Here, we outline the tiling process of a MatMul operator.

\begin{figure}[tbp]
    \centering
    \includegraphics[width=0.9\linewidth]{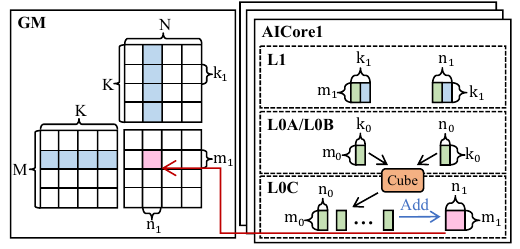}
    \caption{Multi-level tiling decomposition for MatMul on Ascend NPUs.}
    \label{fig:tiling}
\end{figure}

As illustrated in Figure~\ref{fig:tiling} and Table~\ref{tab:notation}, computing an $M \times K$ by $K \times N$ matrix multiplication requires parallelizing the workload across multiple AICores, where each core handles a sub-task of size $m \times k$ and $k \times n$. Within each sub-task, the AICore processes L1 tiles of size $m_1 \times k_1$ and $k_1 \times n_1$, where $m_1\leq m$, $n_1\leq n$, and $k_1\leq k$. These tiles are further partitioned into $m_0 \times k_0$ and $k_0 \times n_0$ micro-tiles for the L0A/L0B buffers and Cube execution.
The corresponding $m_1\times n_1$ FP32 output tile is accumulated in L0C from the $m_0\times n_0$ micro-tiles. We therefore use $m_1n_1\cdot4\text{B}$, rather than the full AICore sub-task size $mn$, to determine whether accumulation remains on chip. If this L0C tile exceeds capacity, the AICore periodically flushes partial sums to Global Memory. Thus, $m,n,k$ determine inter-core task partitioning, whereas $m_1,n_1,k_1$ determine intra-core data reuse and L0C accumulation behavior.

\subsection{Design of the 2D Template Space}\label{sec:template-space}

\begin{table}[tbp]
\centering
\caption{Key notation.}
\label{tab:notation}
\small 
\begin{tabular}{@{}c >{\raggedright\arraybackslash}p{5.8cm} @{}} 
\toprule
\textbf{Symbol} & \textbf{Description} \\ \midrule
$M, N, K$ & Input matrix shapes. \\
$m, n, k$ & AICore sub-task dimensions; they determine inter-core task partitioning. $k=K$ denotes no $K$-split, whereas $k<K$ denotes $K$-axis parallelism. \\
$m_1, n_1, k_1$ & L1 tile dimensions within a sub-task. The $m_1\times n_1$ FP32 result is L0C accumulation tile. \\
$m_0, n_0, k_0$ & L0 micro-tile dimensions for L1$\rightarrow$L0A/L0B transfers and MMAD. \\
$C_{L1}, C_{L0A}, C_{L0B}, C_{L0C}$ & L1, L0A, L0B and L0C buffer capacities, on-chip accumulation requires $m_1n_1\cdot4\text{B}\leq S_{L0C}$ for FP32 results. \\
$P_m, P_n, P_k$ & Task counts along $M, N, K$: $P_m \!=\! \lceil M/m \rceil$, $P_n \!=\! \lceil N/n \rceil$, $P_k \!=\! \lceil K/k \rceil$. \\
\bottomrule
\end{tabular}
\end{table}

When mapping diverse dynamic shapes onto the Ascend NPU, a conflict emerges between inter-core load balancing and intra-core data reuse. Inter-core parallelism is determined by the AICore sub-task dimensions $m,n,k$, whereas GM traffic and L0C occupancy are determined by the L1 tile dimensions $m_1,n_1,k_1$.
Assuming FP16 inputs and no cross-tile residency, each L1 tile reads $m_1k_1+k_1n_1$ elements. Summing over all tiles gives
\begin{equation}
    D_{r} = 2\cdot MNK
    \left( \frac{1}{m_1} + \frac{1}{n_1} \right).
    \label{eq:read-volume}
\end{equation}
Larger $m_1$ and $n_1$ improve data reuse and reduce $D_r$, but the resulting FP32 accumulation tile may exceed L0C capacity. Independently, larger $m$ and $n$ reduce the spatial task count $P_mP_n$ and may leave AICores idle. These two constraints---L0C residency and spatial parallelism---form the axes of our template space.

\begin{figure}[tbp]
    \centering
    \includegraphics[width=\linewidth]{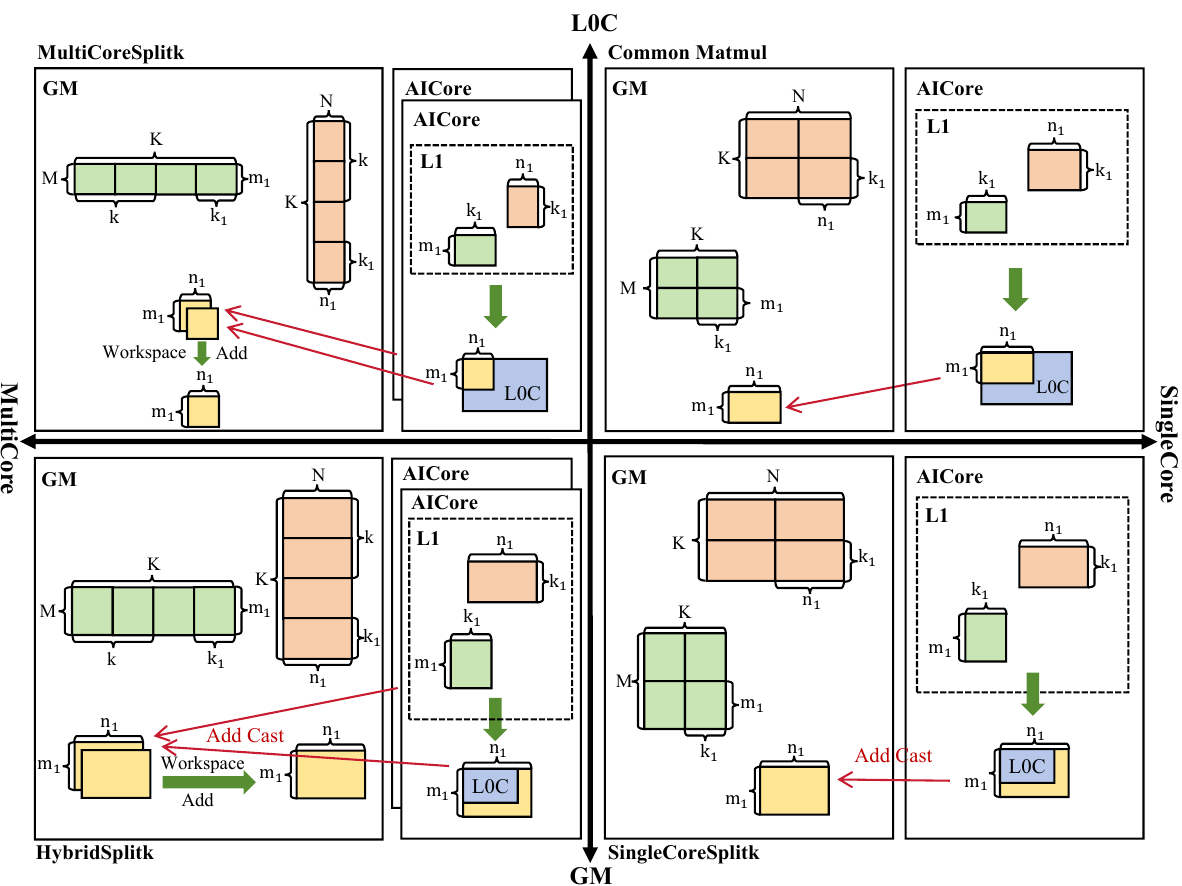}
    \caption{Hardware-aware 2D tiling template taxonomy.}
    \label{fig:template}
\end{figure}

Because no single tiling strategy can reconcile these constraints across all shapes, \sysname organizes four template classes in the hardware-aware 2D space shown in Figure~\ref{fig:template}. The X-axis represents \textit{Spatial Parallelism}, determined by whether the AICore task count is sufficient to saturate the cores. The Y-axis represents the \textit{L0C Accumulation Regime}, determined by whether the FP32 L0C tile fits on chip.

\textbf{Common (Spatial Parallelism, On-Chip Accumulation).} 
This is the baseline template when spatial partitioning generates sufficient tasks to saturate all AICores ($P_m \cdot P_n \geq N_{core}$) and the L0C accumulation tile fits on chip ($m_1 n_1 \cdot 4\text{B} \leq S_{L0C}$). The entire $K$-dimension is processed per core ($k = K$), while each $m_1\times n_1$ output tile is accumulated in L0C. The core performs a single FP32-to-FP16 cast and writes each final output element once to GM, yielding the minimum write volume:
\begin{equation}
    D_{w}^{c} = 2 \cdot MN
\label{eq:dw-common}
\end{equation}

\textbf{SingleCoreSplitK (Spatial Parallelism, Off-Chip Accumulation).} 
For read-bound matrices, \sysname enlarges $m_1$ and $n_1$ to improve L1-level data reuse, even when the resulting $m_1\times n_1$ FP32 tile exceeds L0C capacity. The entire $K$-dimension remains assigned to one core ($k=K$), but the core flushes intermediate partial sums after each $k_1$ slice. This template trades increased write volume for reduced GM read volume:
\begin{equation}
    D_{w}^{s} = 4 \cdot MN \left\lceil \frac{K}{k_1} \right\rceil + 2 \cdot MN
\label{eq:dw-single}
\end{equation}

\textbf{MultiCoreSplitK ($K$-Axis Parallelism, On-Chip Accumulation).} 
For compute-underutilized shapes (e.g., tall-skinny matrices), $P_mP_n<N_{core}$ leaves AICores idle. This template partitions the $K$-dimension across multiple cores ($k<K$), while each local $m_1\times n_1$ FP32 tile still fits in L0C. Every $K$-partition exports one partial result to GM for cross-core reduction, so the write volume depends on the partition count $P_k$:
\begin{equation}
    D_{w}^{m} = 4 \cdot MN \left\lceil \frac{K}{k} \right\rceil + 2 \cdot MN
\label{eq:dw-multi}
\end{equation}

\textbf{HybridSplitK ($K$-Axis Parallelism, Off-Chip Accumulation).} 
This template combines \texttt{SingleCoreSplitK} and \texttt{MultiCoreSplitK} when spatial tasks cannot saturate all AICores ($P_mP_n<N_{core}$) and the enlarged L0C tile exceeds capacity ($m_1n_1\cdot4\text{B}>S_{L0C}$). It partitions the $K$-axis across $P_k>1$ cores while using large $m_1,n_1$ tiles for data reuse. Each AICore consequently flushes local partial sums after every $k_1$ slice, giving:
\begin{equation}
    D_{w}^{h} = P_k \times \left( 4 \cdot MN \cdot \left\lceil \frac{k}{k_1} \right\rceil \right) + 2 \cdot MN = 4 \cdot MN \cdot \left\lceil \frac{K}{k_1} \right\rceil + 2 \cdot MN
\label{eq:dw-hybrid}
\end{equation}
While its mathematical write volume is identical to Equation~\ref{eq:dw-single}, this massive memory penalty is distributed concurrently across $P_k$ AICores.

In summary, these four templates establish a complete design space to balance data transfer volumes and multi-core utilization.
Furthermore, these templates can be combined with other tiling strategies to handle extreme cases. For example, the StreamK strategy can be applied to existing templates to address the tail-iteration load imbalance problem.

\subsection{Mapping Shapes to Templates}
In practice, standard shapes often fall into the \texttt{Common} quadrant. If the enlarged tiles overflow the L0C buffer, execution shifts to \texttt{SingleCoreSplitK}. Conversely, if the shape lacks spatial parallelism, the system triggers \texttt{MultiCoreSplitK} to partition the $K$-axis. Finally, shapes suffering from both core starvation and L0C overflow land in the \texttt{HybridSplitK}.
Here, we provide four representative tiling cases on an Ascend NPU equipped with 16 AICores and a 128 KB L0C. Note that the accumulation on L0C uses the Float32 data type.

\textbf{Case 1: Well-Balanced Workload.}
For $M=N=K=1024$, let each AICore task contain one L1 output tile with $m=m_1=128$ and $n=n_1=256$. The FP32 L0C tile occupies exactly 128 KB, while the spatial partitioning generates $(1024/128)\times(1024/256)=32$ tasks for 16 AICores. The \texttt{Common} template therefore provides both on-chip accumulation and sufficient parallelism.

\textbf{Case 2: Read-Bound Large Matrices.}
For $M=N=4096$ and $K=1024$, increasing the L1 tile from $m_1\times n_1=128\times256$ to $256\times256$ reduces $D_r$ but enlarges the FP32 L0C tile from 128 KB to 256 KB. With $k=K$ and sufficient spatial tasks, this selects \texttt{SingleCoreSplitK}: the larger L1 tile improves reuse, while partial sums are accumulated off chip.

\textbf{Case 3: Compute Underutilization for Skewed Shapes.}
For $M=N=128$ and $K=4096$, let $m=m_1=n=n_1=64$. The 16 KB FP32 L0C tile fits on chip, but the spatial partitioning yields only $(128/64)^2=4$ tasks, leaving 12 of 16 AICores idle. Splitting $K$ across the idle cores selects \texttt{MultiCoreSplitK} and improves parallelism at the cost of cross-core reduction.

\textbf{Case 4: Skewed-Massive Shapes.}
For $M=N=256$ and $K=8192$, let $m=m_1=n=n_1=256$. The configuration produces only one spatial task and a 256 KB FP32 L0C tile. It therefore violates both conditions, selecting \texttt{HybridSplitK} to combine $K$-axis parallelism with off-chip accumulation.

\section{Optimization Space Exploration}\label{sec:optimization}
Although base tiling templates bring substantial performance improvements, achieving peak performance still relies on complex instruction-level optimizations. In this section, we detail how \sysname decouples these optimizations into pluggable components applied to the templates. We also introduce a lightweight cost model to evaluate the benefits of applying the optimizations, guiding developers to choose the optimal operator implementation.


\subsection{Model-Guided Candidate Selection}\label{sec:candidate-selection}
Given an input shape $s=(M,N,K)$, the 2D taxonomy in Section~\ref{sec:template-space} deterministically produces a base implementation $b(s)=(q(s),\boldsymbol{\theta}(s))$, where $q(s)$ is one of the four templates and $\boldsymbol{\theta}(s)=\{m,n,k,m_1,n_1,k_1,m_0,n_0,k_0\}$ contains its associated tiling parameters.
Based on this, \sysname explores a set $\mathcal{S}(b(s))$ of compatible optimization configurations. Each configuration $\mathbf{x}$ specifies both the enabled optimizations and their internal choices, such as the padding format and preloading depth. 
\sysname rejects configurations that violate instruction requirements or any capacity limit in the explicitly managed memory hierarchy. 
Let $\mathbf{M}(b,\mathbf{x},s)$ denote the resulting footprint across the L1 and L0A/B/C buffers, and let $\mathbf{C}_{limit}$ denote their hardware capacities. 
\sysname selects the configuration with the minimum latency predicted by its analytical model $\widehat{T}$:
\begin{equation}
    \mathbf{x}^*(s)=\mathop{\arg\min}_{\mathbf{x}\in\mathcal{S}(b(s))}
    \widehat{T}(b(s),\mathbf{x},s)
    \quad\text{s.t.}\quad
    \mathbf{M}(b(s),\mathbf{x},s)\preceq\mathbf{C}_{limit}.
    \label{eq:goal}
\end{equation}
The exploration is performed offline for the target shape set. The selected optimization configuration, together with its predetermined template and tiling parameters, is encoded into a compact \texttt{TilingKey} and cached with the generated kernel. At runtime, \sysname only maps the incoming shape to its key and retrieves the corresponding kernel.

\subsection{Analytical Performance Modeling}\label{sec:model}
Given the base implementation $b(s)$ and an optimization configuration $\mathbf{x}$, \sysname predicts latency by decomposing execution into a Prologue, a parallel Main Loop, and an Epilogue:
\begin{equation}
\widehat{T}(b,\mathbf{x},s)
= T_{pro} + \max_{j\in\mathcal{P}} T_{main}^{(j)} + T_{epi},
\label{eq:unified_total}
\end{equation}
where $\mathcal{P}$ is the set of participating AICores. The maximum captures the fact that kernel completion is determined by the slowest core.

\textbf{Resource Cost.}
For a transfer stage $r$, its latency on a tile $v$ is modeled as
\begin{equation}
    t_{r,v}=n_{r,v}\lambda_r+
    \frac{D_{r,v}}{BW_r\eta_{r,v}},
    \label{eq:resource_cost}
\end{equation}
where $n_{r,v}$ and $D_{r,v}$ are its instruction count and transferred bytes, $BW_r$ is the peak bandwidth, and $\lambda_r$ and $\eta_{r,v}$ denote the per-instruction overhead and the shape- and alignment-dependent hardware efficiency, respectively. 
We obtain $\lambda_r$ and $\eta_{r,v}$ through empirical offline latency profiling. Cube latency is similarly computed as $t_{cube,v}=N_{mmad,v}\lambda_{mmad}$, where $N_{mmad,v}$ is derived from the L0 tiling parameters. Equation~\ref{eq:resource_cost} evaluates an optimization by propagating its effects on data volume, instruction count, and hardware efficiency.

\textbf{Prologue.}
The Prologue contains optional preprocessing before the kernel main loop. For example, Padding reads an input through MTE2, reformats it on the Vector unit, and writes the transformed data through MTE3. We compute $T_{pro}$ by summing these dependent stages using Equation~\ref{eq:resource_cost}; $T_{pro}=0$ when no preprocessing is required.

\textbf{Main Loop.}
Let $\mathcal{V}_j$ be the set of L1-level tile instances assigned to core $j$. Each tile $v$ produces an output region of at most $m_1\times n_1$ and processes its local $K$ range in
$L_v=\lceil k_v/k_1\rceil$ slices. A slice follows the deterministic MTE2--MTE1--Cube pipeline, with FixPipe additionally appearing as a recurring stage when partial sums must be spilled from L0C.

For a given template and optimization configuration, the steady-state initiation interval is
\begin{equation}
    II_v(b,\mathbf{x})=
    \max_{r\in\mathcal{R}_b} t_{r,v}
    +\Delta_{dep,v}(b,\mathbf{x}),
    \label{eq:initiation_interval}
\end{equation}
where $\mathcal{R}_b$ is the set of recurring hardware stages and $\Delta_{dep,v}$ is the exposed latency caused by data dependencies, synchronization, and buffer reuse hazards. Multi-buffering and preloading reduce $\Delta_{dep,v}$ by enabling overlap across adjacent slices, but are enabled only when the capacity constraint in Equation~\ref{eq:goal} remains satisfied.

Accounting for pipeline fill, steady-state execution, and drain, the main-loop latency is
\begin{equation}
T_{main}^{(j)}=
\sum_{v\in\mathcal{V}_j}
\left[T_{fill,v}+(L_v-1)II_v+T_{drain,v}\right].
\label{eq:main_loop}
\end{equation}
For \texttt{Common} and \texttt{MultiCoreSplitK}, accumulation remains on chip, so FixPipe is excluded from $II_v$ and its final write is included in $T_{drain,v}$. For \texttt{SingleCoreSplitK} and \texttt{HybridSplitK}, L0C overflow causes periodic partial-sum spills, making FixPipe a recurring stage in $\mathcal{R}_b$. Boundary tiles and the last $K$ slice are evaluated using their actual dimensions.

\textbf{Epilogue.}
The Epilogue contains only post-processing required by the selected template and modules:
\begin{equation}
T_{epi}=T_{local\_reduce}+T_{cross\_reduce}
       +T_{format}.
\label{eq:epilogue}
\end{equation}
\texttt{SingleCoreSplitK} requires local reduction of off-chip partial sums, \texttt{MultiCoreSplitK} requires cross-core reduction, and \texttt{HybridSplitK} requires both; the corresponding terms are zero for \texttt{Common}. Format conversion are included only when activated. Each term is evaluated from its data movement and Vector instruction costs using Equation~\ref{eq:resource_cost}.



\subsection{Optimization Library and Modeling}\label{sec:library}
\sysname componentizes common techniques from expert-tuned kernels into the optimization library summarized in Table~\ref{tab:optimizations}. The modules target three bottleneck dimensions. \textit{Bandwidth} modules change transferred bytes $D$ or bandwidth efficiency $\eta$; \textit{pipeline} modules reduce the exposed dependency latency $\Delta_{dep}$; and \textit{compute} modules remove inefficient compute instructions from the critical path. Our analytical model can evaluate their benefits and penalties, both individually and in combination.

\begin{table*}[t]
\centering
\small
\caption{Composable optimization modules and their effects.}
\label{tab:optimizations}
\begin{tabularx}{\textwidth}{@{} l >{\raggedright\arraybackslash}p{2.1cm} >{\raggedright\arraybackslash}X >{\raggedright\arraybackslash}X @{}}
\toprule
\textbf{Dimension} & \textbf{Optimization} & \textbf{Benefit} & \textbf{Cost or constraint} \\ \midrule
\textbf{Bandwidth}
& Padding & Improves MTE2 efficiency through 512B-align. & Adds padded traffic and format conversion. \\ \cmidrule(l){2-4}
& L1 Resident & Reduces GM-to-L1 traffic by reusing the operand. & Increases the L1 footprint. \\ \cmidrule(l){2-4}
& ShuffleK & Improves effective GM bandwidth by staggering inter-core accesses. & May increase TLB pressure; benefit is shape-dependent. \\ \cmidrule(l){2-4}
& Small-$M$ Copy & Reduces conversion and MTE instruction overhead for narrow matrices. & Adds specialized scalar control and applicability constraints. \\ \midrule
\textbf{Pipeline}
& Multi-Buffer & Reduces $\Delta_{dep}$ by alternating independent buffers. & Multiplies the corresponding L1/L0 footprint. \\ \cmidrule(l){2-4}
& Preloading & Reduces $\Delta_{dep}$ and repeated fill cost by advancing the next transfer. & Requires additional L1 stages. \\ \midrule
\textbf{Compute}
& Eliminate Scalar & Reduces control instructions on the critical path. & Specializes the kernel and increases code variants. \\ \bottomrule
\end{tabularx}
\end{table*}

\head{Pipeline Optimization Modeling.}
We use double buffering and preloading to illustrate how the model evaluates pipeline optimizations.
\begin{figure*}[tbp]
    \centering   
    \begin{subfigure}[t]{0.33\linewidth}
        \centering
        \includegraphics[width=\linewidth]{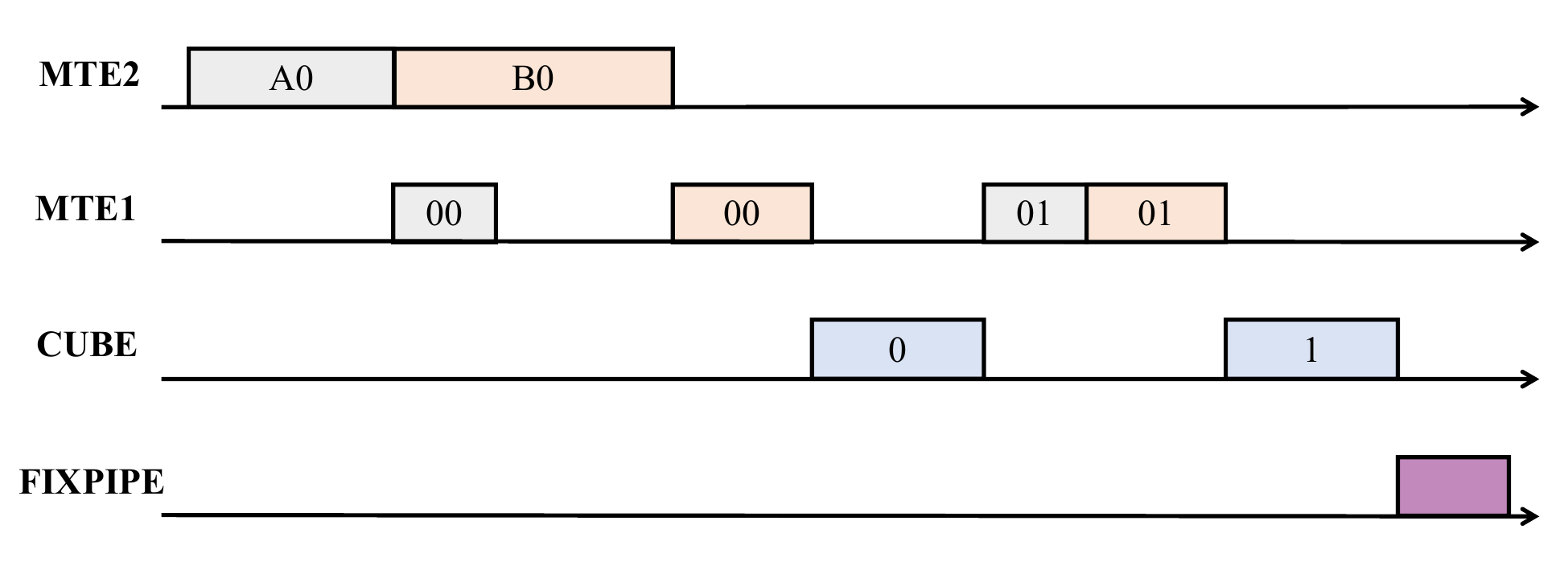}
        \caption{Baseline.}
        \label{fig:preload_baseline}
    \end{subfigure}   
    \hfill  
    \begin{subfigure}[t]{0.33\linewidth}
        \centering
        \includegraphics[width=\linewidth]{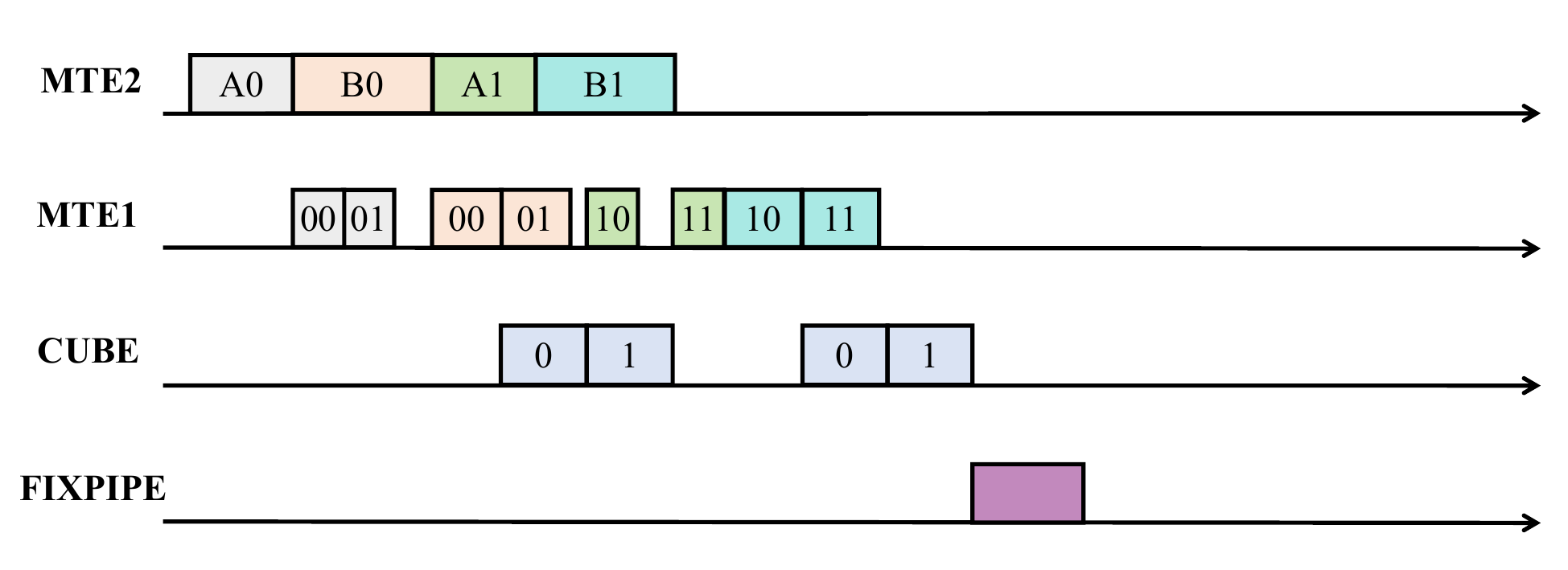}
        \caption{Double-Buffer.}
        \label{fig:preload_db}
    \end{subfigure}
    \hfill 
    \begin{subfigure}[t]{0.33\linewidth}
        \centering
        \includegraphics[width=\linewidth]{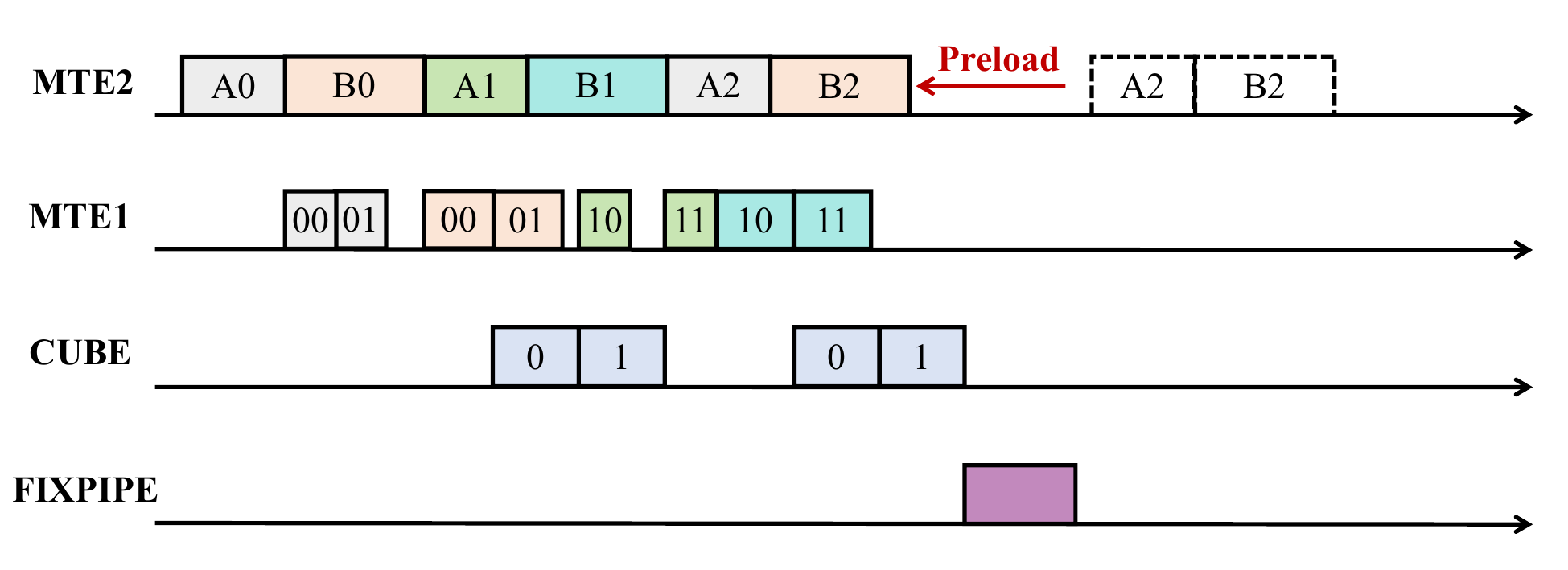}
        \caption{Preloading.}
        \label{fig:preload_preload}
    \end{subfigure}
    
    \caption{Comparison of three pipeline schedules and their effects on buffer-reuse hazards.}
    \label{fig:preload}
\end{figure*}
As shown in Figure~\ref{fig:preload}, these optimizations preserve the transferred bytes and MMAD count, but change the instruction dependencies and buffer footprint. For an L1 tile $v$, let $t_{M,v}=t_{mte2,v}$ denote the MTE2 time of one $K$-slice and let
$t_{P,v}=\max(t_{mte1,v},t_{cube,v})$ denote the overlapped L0/Cube pipeline time. The tile contains $L_v$ slices.

\textit{Baseline.}
As shown in Figure~\ref{fig:preload_baseline}, the same buffer cannot be refilled until its MTE1/Cube consumers finish. MTE2 and the L0/Cube pipeline are serialized across slices, giving
\begin{equation}
T_v^{base}=L_v(t_{M,v}+t_{P,v})+T_{drain,v}.
\label{eq:pipeline_baseline}
\end{equation}

\textit{Double Buffering.}
Figure~\ref{fig:preload_db} alternates two buffer stages in a ping-pong manner. While MTE1/Cube consumes one stage, MTE2 fills the other. After the initial fill, consecutive slices are issued every $\max(t_{M,v},t_{P,v})$ cycles:
\begin{equation}
T_v^{db}=t_{M,v}+(L_v-1)\max(t_{M,v},t_{P,v})
+t_{P,v}+T_{drain,v}.
\label{eq:pipeline_db}
\end{equation}
Thus, the initiation interval in Equation~\ref{eq:initiation_interval} becomes $II_v^{base}=t_{M,v}+t_{P,v}$ for the baseline and $II_v^{db}=\max(t_{M,v},t_{P,v})$ for double buffering.

\textit{Preloading.}
Double buffering still pays an MTE2 fill at the beginning of every L1 tile. Figure~\ref{fig:preload_preload} advances the first MTE2 transfer of tile $v+1$ into the L0/Cube pipeline of tile $v$. For $Q_j$ regular tiles assigned to core $j$, each containing $L$ slices with stage times $t_M$ and $t_P$, the continuous pipeline is modeled as
\begin{equation}
T_{main,reg}^{(j),pre}=t_M+(Q_jL-1)\max(t_M,t_P)
+t_P+Q_jT_{drain}.
\label{eq:pipeline_preload}
\end{equation}
In contrast, double buffering executes the same tiles in $Q_j[t_M+(L-1)\max(t_M,t_P)+t_P+T_{drain}]$.

The model substitutes Equations~\ref{eq:pipeline_baseline}--\ref{eq:pipeline_preload} into the latency model and selects the fastest legal schedule. Double buffering or preloading is rejected if it violate hardware limits. The remaining optimizations in Table~\ref{tab:optimizations} are evaluated analogously by updating their affected model variables.

\section{Evaluation}\label{sec:evaluation}
In this section, we evaluate the performance of the MatMul operators generated by \sysname against baselines on Ascend. We also conduct comprehensive scalability and ablation studies to demonstrate \sysname's effectiveness. 


\subsection{Experimental Setup}

\head{Hardware.} The evaluation of our MatMul templates was conducted on Linux servers equipped with Huawei Ascend 910B NPUs. Each NPU card accommodates 24 AICores, delivering a theoretical peak FP16 compute throughput of 376 TFLOPS and an off-chip Global Memory (GM) to L1 Cache bandwidth of approximately 1.6 TB/s. 

\head{Software.} The detailed software environment included CANN 9.0.T500, Python 3.13.11, CMake 3.31.10, Clang 15.0.5, and GCC 10.3.1. We utilized the \texttt{msprof} profiling tool included in the CANN package to collect hardware metrics and execution traces during operator runtime.

\head{Baselines.} We compared \sysname's MatMul operator implementations with those provided by the Ascend vendor library (ACLNN~\cite{cann_website}) and the basic CATLASS templates~\cite{catlass_gitcode}.
Given high auto-tuning overhead and the lack of awareness regarding the Ascend memory architecture, the latency of Triton-Ascend~\cite{triton_ascend} operators is currently far higher than that of the aforementioned options. Therefore, it is not being considered.
By comparing the practical execution times of these implementations, we demonstrate the generalization capability and effectiveness of \sysname.

\head{Workloads.} To comprehensively assess the effectiveness, we evaluated the single-operator latency and end-to-end latency of real-world recommendation scenarios, which are heavily affected by dynamic input shapes. 
For single-operator latency, we construct a large-scale dataset of roughly 80,000 diverse and dynamic MatMul input shapes derived from real-world industrial input ranges and distributions. For end-to-end performance, we benchmark inference workloads from representative recommendation models, including MMOE~\cite{yu2024mmoeenhancingmultimodalmodels}, DLRM~\cite{naumov2019deeplearningrecommendationmodel}, DCN V2~\cite{10.1145/3442381.3450078}, ESMM~\cite{ma2018entirespacemultitaskmodel}, and RankMixer~\cite{10.1145/3746252.3761507}.

\begin{figure*}[tbp]
     \centering
     \begin{minipage}[b]{0.32\linewidth}
        \centering
        \includegraphics[width=\linewidth]{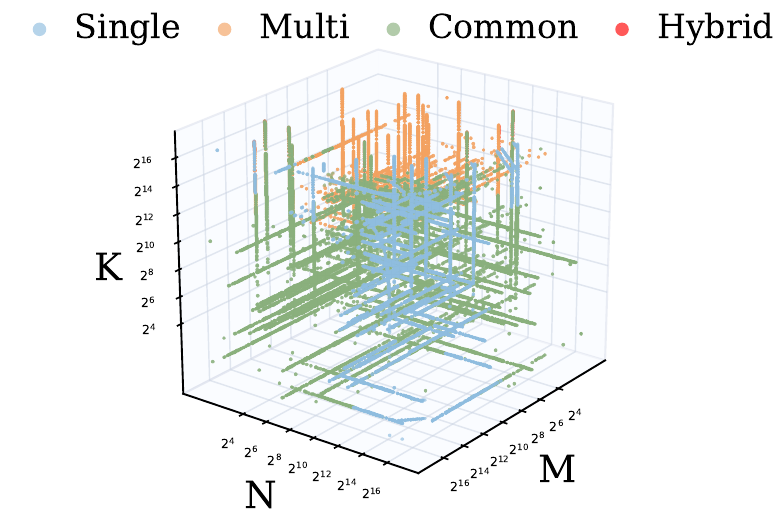}
        \caption{Statistical distribution of input shapes.}
        \label{fig:mnk_shapes}
     \end{minipage}
     \hfill
     \begin{minipage}[b]{0.32\linewidth}
        \centering
        \includegraphics[width=\linewidth]{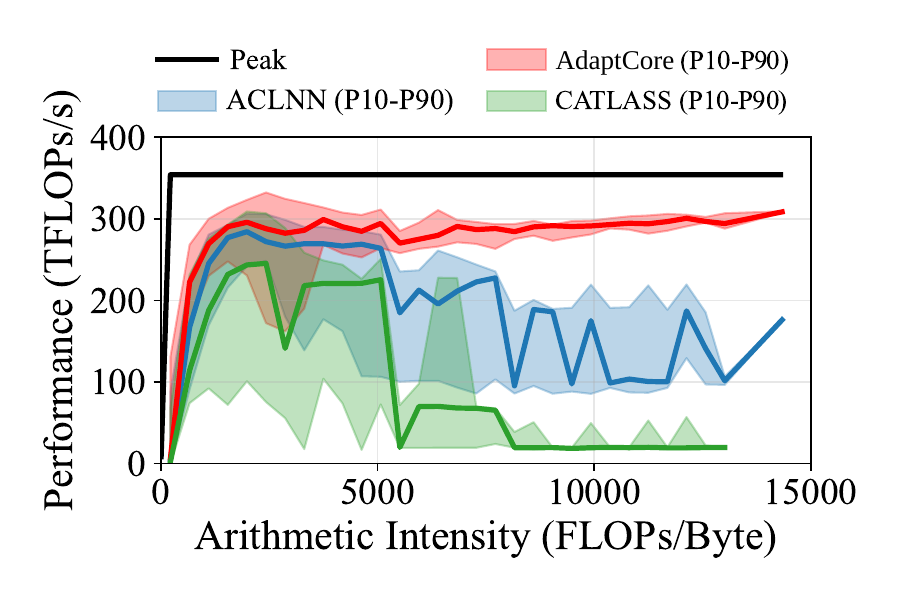}
        \caption{Roofline model of different Matmul.}
        \label{fig:roofline}
     \end{minipage}
     \hfill
     \begin{minipage}[b]{0.32\linewidth}
        \centering
        \includegraphics[width=\linewidth]{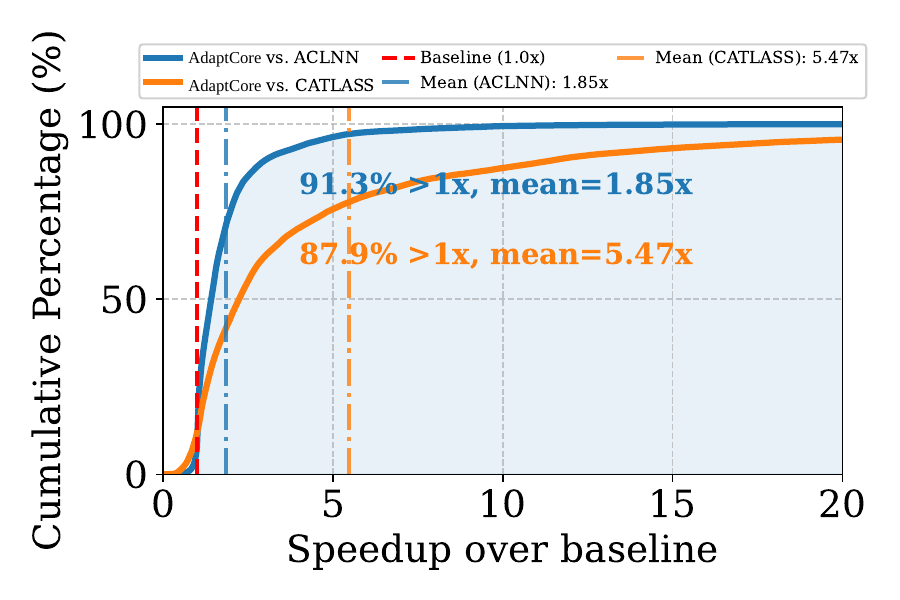}
        \caption{CDF of speedup.}
        \label{fig:cdf}
     \end{minipage}
\end{figure*}

\subsection{Single-Operator Latency}
Figure~\ref{fig:mnk_shapes} visualizes the distribution of approximately 80,000 input shapes in the 3D $(M, N, K)$ space with their assigned tiling templates. The dataset spans regular to extreme shapes, fully capturing the variance of dynamic workloads. Through its hardware-aware taxonomy, \sysname maps these shapes to distinct templates, revealing clear shape characteristics. For example, shapes with a large $K$ dimension favor the \texttt{MultiCoreSplitK} template to maximize multi-core parallelism, while shapes with larger $M$ and $N$ dimensions tend to use \texttt{SingleCoreSplitK} to optimize memory bandwidth. Finally, due to its strict triggering conditions, the \texttt{HybridSplitK} template is rarely activated, appearing only 591 times.

We compare the operator execution latency of three implementations: the ACLNN library, basic CATLASS template library,
and \sysname. Unlike the hand-tuned ACLNN implementation, basic CatLASS exposes explicit tiling primitives but requires the user to manually select the template and schedule the pipeline. 

Figure~\ref{fig:roofline} quantifies the hardware utilization of different operator implementations for these dynamic shapes. The Roofline-style plot summarizes the performance distribution of four GEMM implementations. The median performance (solid lines) and P10–P90 interval (shaded areas) are computed via arithmetic intensity binning, enabling a robust comparison of both peak throughput and performance variability.
Relying on static tiling rules, the baseline ACLNN operator severely underutilizes both compute and memory bandwidth when encountering irregular shapes (blue), achieving an average throughput of only 71.50 TFLOPs/s. In contrast, \sysname significantly boosts both arithmetic and memory bandwidth utilization (red), driving the average throughput up to 90.96 TFLOPs/s. Meanwhile, the basic CATLASS matmul (green) achieves an average throughput of 55.81 TFLOPs/s on dynamic shapes, 
Both the baselines fall considerably behind \sysname. 

The CDF of latency-speedup in Figure~\ref{fig:cdf} also confirms the trend: \sysname achieves a mean speedup of 1.85$\times$ over ACLNN, and 5.47$\times$ over basic CATLASS
The largest gains occur on highly skewed shapes, for which generic template and pipeline choices leave substantial hardware resources underutilized.




\begin{figure*}[tbp]
     \centering
     \begin{minipage}[b]{0.32\linewidth}
        \centering
        \includegraphics[width=\linewidth]{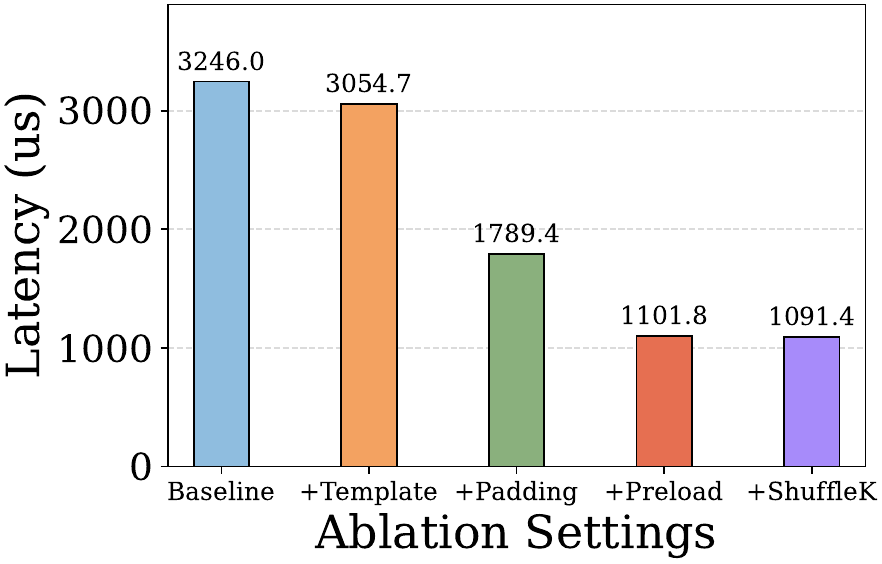}
        \caption{Ablation analysis on a representative skewed shape.}
        \label{fig:ablation}
     \end{minipage}
     \begin{minipage}[b]{0.32\linewidth}
        \centering
        \includegraphics[width=\linewidth]{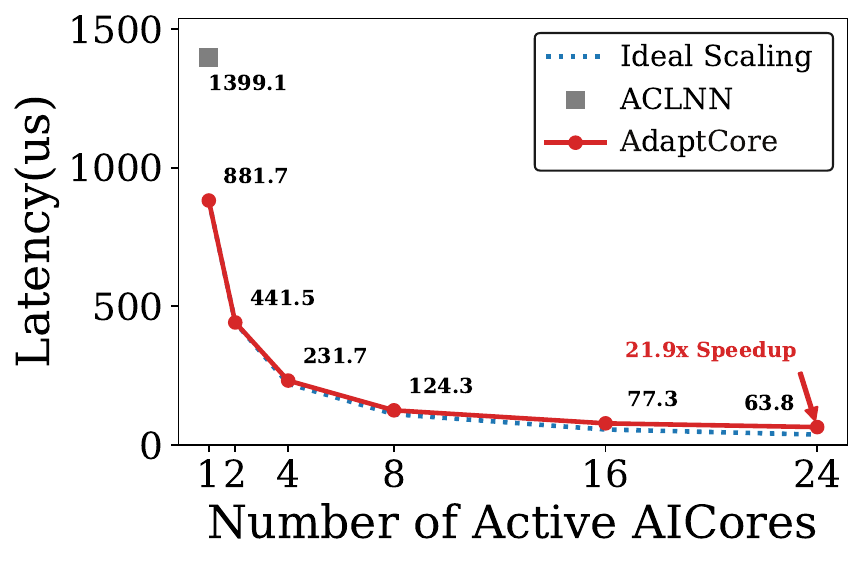}
        \caption{Multi-core scalability analysis.}
        \label{fig:scalability}
     \end{minipage}
     \hfill
     \begin{minipage}[b]{0.32\linewidth}
        \centering
        \includegraphics[width=\linewidth]{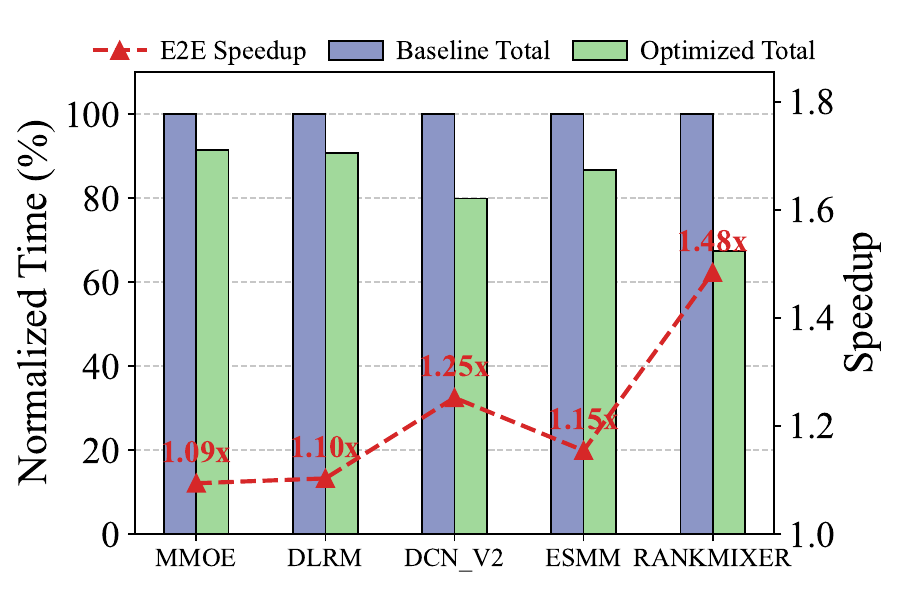}
        \caption{Latency reduction of end-to-end latency.}
        \label{fig:end_to_end}
     \end{minipage}
\end{figure*}

\subsection{Ablation Study}
To evaluate the performance gains from \sysname's tiling templates and composable optimizations, we conduct an ablation study on a representative regular shape ($M=12288$, $N=688$, $K=16991$). As illustrated in Figure~\ref{fig:ablation}, using the native ACLNN implementation (3246.0 $\mu$s) as the baseline, we incrementally activate optimizations. First, applying the \textit{Common Template} slightly reduces latency to 3054.7 $\mu$s by automatically deriving optimal tiling parameters for this uniform shape. Next, integrating \textit{Memory Padding} drastically cuts the latency to 1789.4 $\mu$s. This optimization uses the Vector unit to pre-transform the right matrix into the NZ format and aligns the stride to 512B, strictly matching the L1 buffer layout. This eliminates the massive bandwidth degradation caused by unaligned memory accesses and ND-to-NZ format conversions. Building on this, we enable \textit{Preloading} ($p=1$) to perfectly overlap Cube computation with GM to L1 data fetches. This pipelining effectively hides the memory latency, further dropping the execution time to 1101.8 $\mu$s. Finally, we apply \textit{ShuffleK}, which temporally staggers the data access sequence to prevent concurrent read conflicts across multiple AICores. However, since the read bandwidth is already near-optimal due to previous steps, the latency yields a marginal reduction to 1091.4 $\mu$s. Overall, these results strongly validate the effectiveness of \sysname's tiling templates and composable optimizations.

\subsection{Scalability Analysis}
To evaluate \sysname's impact on multi-core parallelism efficiency, we conduct a comprehensive scalability study. We select a highly skewed shape with a large $K$ dimension ($M=3, N=256, K=87087$) and scale the number of available AICores on the Ascend NPU from 1 to 24 to monitor latency variations.
As illustrated in Figure~\ref{fig:scalability}, the native ACLNN implementation suffers from a severe scalability bottleneck. Constrained by static tiling rules, the total number of parallel tasks in ACLNN is strictly limited by the minuscule $M$ and $N$ dimensions. Consequently, it activates only a single core (yielding a latency of 1399.1 $\mu$s) and leaves the remaining 23 AI Cores completely idle, failing to extract any scaling benefits. 
In contrast, \sysname leverages its hardware-aware taxonomy to dispatch the \texttt{MultiCoreSplitK} template. By partitioning the massive $K$ axis and coordinating inter-core reductions, \sysname achieves exceptional load balancing and sustained latency reductions. Its execution time scales near-ideally, plunging from 881.7 $\mu$s on a single core down to 63.8 $\mu$s on 24 cores. At full 24-core utilization, \sysname delivers a massive 21.9$\times$ speedup over the native baseline.

\subsection{End-to-End Latency}
To evaluate the end-to-end benefits of \sysname in real-world scenarios, we benchmark the inference latency across representative recommendation models (e.g., MMoE, DLRM, DCN V2, ESMM, and RankMixer). By seamlessly replacing the native General Matrix Multiply (GEMM) operators (including standard MatMul, AddMM, GroupMM, and BatchMM) with \sysname’s implementation, we observe substantial speedups in end-to-end time.
As illustrated in Figure~\ref{fig:end_to_end}, we present the normalized execution time breakdown and the corresponding speedups. In the ACLNN implementation of different models, GEMM operators are the primary bottleneck. Following our optimization, the latency of these GEMM operators is drastically compressed while the non-GEMM operators remain constant. Eliminating the GEMM bottleneck yields the 1.09$\times$-1.48$\times$ speedups of end-to-end latency.


\subsection{Performance Model} 
We first evaluate the prediction accuracy of the analytical model. Because the model is designed to rank candidate configurations rather than predict cycle-accurate absolute latency, moderate point-wise error is expected. 
On the 80{,}000-shape benchmark, we exhaustively measure all valid template--optimization configurations and treat the fastest as the oracle. \sysname selects the exact oracle configuration for 63.86\% of the shapes. Importantly, the selected kernels still outperform ACLNN on 91.3\% of the shapes, confirming that the model-guided ranking is effective in practice. 

We then measure the deployment cost of the cost model. The model is invoked only at template selection time, before operator kernel generation. Its main overhead comes from evaluating the performance formulas for each candidate strategy and checking capacity constraints. The average modeling time per shape is on the order of 10 milliseconds. Since this process can be performed offline, its cost remains acceptable even for a dataset of 80,000 shapes. This confirms that \sysname's analytical model is practical for deployment.
\section{Limitations and Future Work}\label{sec:limitations}
We list the current limitations of \sysname and outline several directions for future exploration.

\textit{Architectural Generality.} 
While the methodology of \sysname (hardware-aware tiling taxonomy and optimization modeling) is theoretically generalizable, its specific templates and empirical estimates heavily rely on Ascend hardware. Porting \sysname to other Domain-Specific Architectures (DSAs) with different memory hierarchies would require redefining the taxonomy and analytical performance models.

\textit{Operator Scope and Workload Shifts.} 
Currently, \sysname focuses on matrix multiplication operators (e.g., MatMul, GroupMM, BatchMM). As AI workloads evolve towards ultra-low precision (FP8/INT4), highly unstructured sparsity, or complex operator fusion, execution bottlenecks may shift. Moving forward, we plan to extend our methodology to a broader spectrum of complex operators (e.g., FlashAttention, Convolution) by formalizing their hardware constraints.

\textit{Agent-Driven Operator Generation.} 
Recent research has explored the use of LLM agent workflows (e.g., CUDA Agent~\cite{dai2026cudaagentlargescaleagentic}, TensorCoder~\cite{zhang2020tensorcoderdimensionwiseattentiontensor}) for auto kernel generation. However, due to a lack of hardware knowledge of Ascend NPUs and reliable training data, these models perform poorly on Ascend. In the future, we aim to integrate our architectural insights into an Ascend operator agent to assist efficient bottleneck localization, bug elimination, and performance tuning.

\section{Conclusion}
\label{sec:conclusion}
In this paper, we propose \sysname to resolve the severe dynamic shape crisis on explicitly decoupled Ascend NPUs. By introducing a hardware-aware 2D tiling taxonomy, an analytical performance model and a composable optimization library, \sysname mathematically navigates strict physical constraints to achieve $O(1)$ optimal runtime dispatching. Comprehensive evaluations across 80,000 industrial shapes and representative end-to-end models demonstrate that \sysname delivers a remarkable $1.85\times$ mean speedup over native vendor libraries. Ultimately, \sysname successfully bridges deep architectural awareness with runtime flexibility, establishing a new paradigm for universally high-performance AI execution on explicitly controlled hardware.

\section*{Acknowledgment}
The authors acknowledge OpenAI GPT 5.6 Sol as a language assistant to polish English prose and improve overall readability. The authors carefully reviewed all AI-assisted edits and take full responsibility for the content of this paper.

\bibliographystyle{ACM-Reference-Format}
\bibliography{reference}

\end{document}